\documentclass{eptcs}
\providecommand{\event}{ADG 2021} % Name of the event you are submitting
\usepackage[utf8]{inputenc}
\usepackage{amssymb}
\usepackage{numprint}
\usepackage[english]{babel}
\usepackage{fontenc}
\usepackage{graphicx}
\usepackage{hyperref}
\usepackage{fancyvrb}
\usepackage{xcolor}
\usepackage{cite}
\usepackage{enumitem}
\usepackage[ruled,linesnumbered,resetcount]{algorithm2e}
\hypersetup{
    colorlinks,
    linkcolor={red!50!black},
    citecolor={blue!50!black},
    urlcolor={blue!80!black}
}

\DefineVerbatimEnvironment%
{CoqVerbatim}{Verbatim}
{frame=lines, fontsize=\small}

\newtheorem{theorem}{Theorem}
\newtheorem{lemma}[theorem]{Lemma}
\newtheorem{property}[theorem]{Property}
\newcommand{\rk}{\mbox{rk}}

\newcommand{\Point}{\mbox{Point}}
\newcommand{\point}{\mbox{Point}}

\newcommand{\mline}{\mbox{Line}}
\newcommand{\si}{if}
\newcommand{\alors}{then}
\newcommand{\et}{and }

\def\orcidID#1{\unskip$^{\mbox{\href{https://orcid.org/#1}{\scriptsize{[#1]}} }}$}

\title{Mechanization of Incidence Projective Geometry in 
Higher Dimensions, a Combinatorial Approach}

\author{Pascal Schreck\orcidID{0000-0003-0199-0598}
\institute{ICube, UMR 7357 CNRS \\ Université de Strasbourg, France}
\email{schreck@unistra.fr}
\and
Nicolas Magaud\orcidID{0000-0002-9477-4394} 
\institute{ICube, UMR 7357 CNRS \\ Université de Strasbourg, France}
\email{magaud@unistra.fr}
\and
David Braun\orcidID{0000-0002-0839-0934}
\institute{ICube, UMR 7357 CNRS \\ Université de Strasbourg, France}
\email{dmp.braun@gmail.com}
}

\def\titlerunning{Mechanization of Incidence Projective Geometry in 
Higher Dimensions, a Combinatorial Approach}
\def\authorrunning{Schreck, Magaud \& Braun}

\begin{document}

\maketitle

\begin{abstract}
 Several tools have been developed to enhance automation of theorem
 proving in the 2D plane. However, in 3D, only a few approaches have
 been studied, and to our knowledge, nothing has been done in higher
 dimensions. In this paper, we present a few examples of incidence
 geometry theorems in dimensions 3, 4, and 5. We then prove them with the help of a combinatorial prover based on matroid theory applied to geometry.
%  A lot have been done in automation of theorem proving in the 2D plane. In 3D however, only a few approaches have been studied, and to our knowledge nothing has been done in higher dimensions. In this paper, we present few examples of incidence geometry theorem in dimension 3, 4 and 5 proved with the help of a combinatorial prover based on matroid theory applied to geometry.
\end{abstract}

%\smallskip
%\noindent \textbf{Keywords.} mechanization of incidence geometry, matroid

\section{Introduction}

Most of the studies about automation of geometric reasoning are about plane geometry in dimension 2 (See for instance \cite{Gelernter,chou,chou94,YeChouGao,Janicic,Botana}) and there are few studies in 3D  (See for instance \cite{chou, Roanes-MaciasR06, Affeldt,DBLP:conf/adg/FuchsT10}). Besides, most of these works refer to rich geometries, that is geometries with some metric features like area, angle, and, more often, coordinates. In this paper, we present elements of automation for a very simple geometry but in higher dimensions.

The geometric framework we choose is the simplest possible and corresponds to the very first axioms of Hilbert's formalization of geometry: only points, lines and their incidence relationship are considered. Note also that the notion of incidence is weaker than Tarski's betweenness where order between points is involved. Informally, the plane incidence projective geometry can be mainly described by two very simple properties which are to be fulfilled: two different points define a line, two different lines define a point. We will see later that the description is a bit more complicated when it comes to precisely defining a higher dimension.

In this paper, we describe a method based on the theory of matroid allowing to mechanize some parts of a proof in incidence geometry in a homogeneous way whatever the dimension. As often, the axioms where existential quantification introduces new points are difficult to handle and, in our method, they are left to the human geometer. We design and implement an automatic prover that initially aimed to prove short parts of proofs, but also succeeded in proving difficult incidence geometry theorems in dimension 3. We present this prover and some experiments in dimension 3, 4 and 5.

The rest of the paper is organized as follows. In Section~\ref{sect:geometry}, we present the synthetic axiom system for projective
incidence geometry . In Section~\ref{sect:matroid}, we introduce the
concept of rank and show how to use it to carry out a proof in
geometry.  In Section~\ref{sect:Bip}, we describe our automatic
prover. In Section~\ref{sect:examples}, we present several examples of
application of our prover, ranging from 3D configurations to 5D ones.
In Section~\ref{sect:conclusion}, we make some concluding remarks.

% -----------my outline
% The rest of the paper is organized as follows. In section \ref{sect:geometry}, we give more details on sets of axioms related to incidence projective geometry. In section \ref{sect:matroid}, we present the matroidal point of view on incidence geometry and we explain in section \ref{sect:Bip} how it has been used to implement an automatic prover so-called Bip. We present then, in section \ref{sect:examples}, some examples in various dimensions.
%TODO

%---------------------------------------------------------------------------------
%
%      Incidence projective geometry
%
%---------------------------------------------------------------------------------

\section{Incidence projective geometry}\label{sect:geometry}

Incidence geometry is one of the simplest geometric theories where only points and lines as well as an incidence relationship between them are considered. This geometric framework is even made simpler when considering its projective version where no axiom of parallelism is used.
For instance, the notion of incidence projective plane is simply defined by imposing that two different points define a line, two different lines define a point, each line is incident to at least three points and there are two lines. Formally, this can be stated by the following set of axioms where the incidence relation is denoted by $\in$ in infix notation:
\begin{enumerate}[label=($\roman*$)]
 \item $\forall\ A\ B : \point\ \exists\ d : \mline, A \in d \wedge B \in d$
 \item $\forall\ d\ d'  : \mline\ \exists\ A : \point, A \in d \wedge A \in d'$
 \item $\forall\ A\ B : \point\ \forall\ d\ d' : \mline, A \in d \wedge B \in d \wedge A \in d' \wedge B \in d' \Rightarrow A =B \vee d = d'$
 \item $\forall\ d : \mline\ \exists\ A\ B\ C : \point, A \neq B \wedge A \neq B \wedge B \neq C \wedge A \in d \wedge  B \in d \wedge C \in d$
 \item $\exists\ d\ d' : \mline, d \neq d'$
\end{enumerate}

This elegant formulation is proper to the plane. It puts into light the symmetry between points and lines but it hides the fact that axioms ($i$), ($iii$) and ($iv$) are general whereas axioms ($ii$) and ($v$) are related to the plane. When no mention of dimension is required, axiom ($ii$) is just replaced by Pasch axiom stating that if two lines $d_1$ and $d_2$ cut each other, then every pair of lines $d_3$ and $d_4$ intersecting $d_1$ and $d_2$, also cut each other. Using just the incidence relation leads to a bit more complicated formulation:\\
$
\begin{array}{l}
 (vi)\ \forall A\ B\ C\ D\ M\ : \point\ \forall\ d_1\ d_2\ d_3\ d_4\ : \mline, \\
     \mbox{\hspace{0.7cm}} A \in d_1\ \wedge B \in d_1\ \wedge\ M \in d_1\ \wedge\ C \in d_2\ \wedge\ D \in d_2\ \wedge\ M \in d_2\ \wedge\\
     \mbox{\hspace{0.7cm}} A \in d_3\ \wedge\ C \in d_3\ \wedge\ B \in d_4\ \wedge\ D \in d_4\ \Rightarrow\\
     \mbox{\hspace{0.7cm}} \exists\ P : \point, P \in d_3\ \wedge\ P \in d_4
\end{array}
$

We call general axioms of incidence projective geometry, the four axioms ($i$), ($iii$), ($iv$) and ($vi$). Axiom $(ii)$ is called the upper dimension axiom for the plane since it defines an upper  bound: there is \emph{not enough space} for two lines not cutting each other, the dimension (here in its intuitive meaning) is at most 2. The axiom ($v$) is called the lower dimension axiom for the plane. It defines a lower bound: the geometric framework is not reduced to a single line.

In higher dimensions, the general axioms are kept and other upper and lower dimension axioms are considered. For instance, in dimension 3, we get:
\begin{itemize}
 \item[$(vii)$] $\exists\ d\ d' : \mline, \neg( \exists A : \point, A \in d\ \wedge\ A \in d')$
 \item[$(viii)$]
 $\forall\ d\ d'\ d'' : \mline\ \exists\ A\ B\ C : \point\ \exists\ \delta : \mline,   \\
   A \in d \wedge A \in \delta \wedge
   B \in d' \wedge B \in \delta \wedge
   C \in d'' \wedge C \in \delta
 $
\end{itemize}
Axiom $(vii)$ defines the lower bound: it is the negation of axiom $(ii)$ (upper dimension for the plane) an axiom $(viii)$ is the upper dimension axiom.% and line $\delta$ is called the transversal lines for $d, d'$ and $d''$. 

An upper dimension axiom for dimension 4 can be stated by saying that the intersection of 3D spaces is at least a plane, or, with a formula:
\begin{itemize}
 \item[$(ix)$] 
 $\forall\ d_1\ d_2\ d_3\ d_4 : \mline\ \exists\ A\ B\ C\ D\ P : \point\ \exists\ \delta\ \delta' : \mline,\\
 P \in \delta \wedge A \in \delta \wedge B \in \delta \wedge A \in d_1 \wedge B \in d_2\\
 P \in \delta' \wedge C \in \delta' \wedge D \in \delta' \wedge C \in d_3 \wedge D \in d_4
 $
\end{itemize}

It is important to recall that no coordinates are involved in incidence geometry \footnote{Indeed, Pappus's property must hold in order to define algebraic structures such as a field on an incidence space.}. This geometry being much more general than the usual ones, a lot of familiar theorems do not hold. This is why we distinguish between property and theorem to avoid ambiguities. For instance, Pappus's property does not hold in general in an incident projective space. For the sake of completeness, let us recall some famous properties usually linked to projective geometry.

%\textbf{Deasrgues's property.}
\begin{property}[Desargues]\label{thm:desargues}
 Let $ABC$ and $A'B'C'$ be two triangles in perspective from $O$ in an incidence projective plane, that is the respective triple $(O,A, A')$, $(O,B, B')$ and $(O,C,C')$ are collinear and such that the corresponding edges $AB$ and $A'B'$ intersect in point $\gamma$, $AC$ and $A'C'$ in point $\beta$ and, $BC$ and $B'C'$ in point $\gamma$; then points $\alpha, \beta$ and $\gamma$ are collinear. 
\end{property}

This fundamental property is linked to a possible algebraic structure of incidence geometry: this property in an incidence plane holds if and only if a division ring can be defined from the geometry. Furthermore, if the plane is embedded in a 3D space, this property becomes a theorem. 
%With Burnside's theorem, we get that all finite incidence spaces with dimension greater than 2 come from a Galois's field. In section \ref{}, we prove that this theorem holds in dimension greater than 2.
Then, we have a stronger property:

\begin{property}[Pappus]
 Let $a$ and $e$ be two distinct lines and let
 $A$, $B$, $C$, $A'$, $B'$ and $C'$ be six different points with $A$, $B$ and
 $C$ belonging to $a$ and $A'$, $B'$ and $C'$ belonging to $e$. These
 points define respectively the lines $l_{AB'}, l_{A'B}, l_{AC'}, l_{A'C},$
 $l_{BC'}, l_{B'C}$. The three intersection points   $X=l_{AB'} \cap l_{A'B}$ and $Y=l_{AC'}\cap l_{A'C}$ and $Z=l_{BC'} \cap l_{B'C}$ are collinear.
\end{property}

The next property is a pure 3D product. 
\begin{property}[Dandelin-Gallucci]
 In a 3D incidence projective space, three skew lines $a$, $b$ and $c$ being given and three other skew lines $e$, $f$ and $g$ being also given, such that every line in $\{a, b, c\}$ meets every line in $\{e, f, g\}$. Then, all pairs of lines $d$ and $h$ such that $d$ meets lines $e$, $f$ and $g$ and $h$ meets every line $a$, $b$ and $c$, are concurrent.
\end{property}

%---------------------------------------------------------------------------------
%
%       A matroid implementation of incidence projective geometry
%
%---------------------------------------------------------------------------------

\section{A matroid implementation of incidence projective geometry}\label{sect:matroid}

Matroid theory was introduced by Whitney in 1935 to abstractly capture the essence of dependence of the columns of matrices. A presentation of matroid theory can be found in \cite{oxley}. Applications of that theory can be found in geometry (see \cite{MS06} for instance) but also in different computer science domains such as algorithmic and graph theory \cite{edmond,tutte}.

Where a certain notion of (in)dependence exists, a matroid structure is hidden. It is the case of the incidence projective geometry where 3 points are dependent if they are collinear, 4 points are dependent if  any two lines defined by two of them intersect each other, and so on. Such a dependence relationship is fully defined through new notions like flats spanned by some points, closure and rank function. We will not go further in that direction, it is enough to state that in the one hand, our projective rank function is precisely a rank function related to a matroid, and, on the other hand, the rank of a set of points $S$ intuitively corresponds to the dimension of the affine set $E$ spanned by $S$ plus one: $\rk(S)=\dim(E)+1$. The important result is that all the axioms and all the properties given above can be expressed through that rank function (See \cite{desargueslong, Braun2019}). 
Using rank function as defined above allows to simply
captures incidence, collinearity and coplanarity properties.
It makes then proof automation easier because it does not directly handle  
lines, planes or other spaces.   

Formally, an integer function \emph{rk} on the powerset of a set $E$ is the rank
function of a matroid if and only if conditions of Table \ref{rfp} are
satisfied. %Here again, we consider a classical logic framework.

\begin{table}[h]
\begin{itemize}
\item[$(A_1)$] 
\mbox{\hspace{0.2cm}}$\forall\ X,\ 0 \leq \rk(X) \leq \big| X\big|$
\item[$(A_2)$]  
\mbox{\hspace{0.2cm}}$\forall\ X\ Y,\ X \subseteq Y \Rightarrow \rk(X) \leq \rk(Y)$
\item[$(A_3)$] 
\mbox{\hspace{0.2cm}}$\forall\ X\ Y,\ \rk(X\cup Y) + \rk(X\cap Y) \leq \rk(X) + \rk(Y)$ 
\end{itemize}
\caption[]{Rank function properties, $X$ and $Y$ are finite subsets of $E$ - Axiom $A_3$ expresses the sub-modularity of \rk}\label{rfp}
\end{table}

\begin{table}[h]
\begin{tabular}{ll}
$\rk(\{A,B\})$ = 1 & ~~~~~$A = B$\\
$\rk(\{A,B\}) = 2$ & ~~~~~$A \neq B$\\
$\rk(\{A,B,C\}) \leq 2$ & ~~~~~points $A,B,C$ are collinear\\
$\rk(\{A,B,C\}) = 3$ & ~~~~~points $A,B,C$ define a plane\\
$\rk(\{A,B,C,D\}) = 4$ & ~~~~~points $A,B,C,D$ are not coplanar\\\\
\end{tabular}
\caption[]{Ranks of some sets of points and their geometric interpretations}\label{rs}
\end{table}

Using a rank function, it can be shown that every projective space has 
a matroid structure, but the converse is not true. To
capture the geometric aspects, we add axioms formally stated on Table~\ref{rps}.  Note that we can have a simpler formulation for the lower and upper dimension axioms $(A_8), (A_9)$ and $(A_{10})$. Axioms $(A_8)$ and $(A_9)$ only express the upper bound of the rank function in 3D. Axiom $(A_{10})$ does not translate axiom $(viii)$, it simply expresses that, in 3D, a line and a plane always intersect each other. 

\begin{table}[h]
\begin{itemize}
\item[$(A_4)$] 
$\forall P : \Point,\ \rk(\{P\}) = 1$
\item[$(A_5)$] 
$\forall P\ Q: \Point, P \neq Q \Rightarrow \rk(\{P, Q\}) = 2$
\item[$(A_6)$] 
$\forall\ A\ B\ C\ D: \Point,\ \rk(\{A, B, C, D\}) \leq 3~ \Rightarrow $\\
$\exists\ J: \Point,\ \rk(\{A, B, J\}) = \rk(\{C, D, J\}) = 2$
\item[$(A_7)$] 
$\forall\ A\ B: \Point,\ \exists\ C: \Point,$\\ $\rk(\{A, B, C\}) = \rk(\{B, C\}) = \rk(\{A, C\}) = 2$
\item[$(A_8)$] 
$\exists A\ B\ C\ D : \Point,\ \rk(\{A, B, C, D\}) \geq 4$
%\rule{330pt}{0.4pt}
%\vspace*{0.1cm}
\item[$(A_9)$] 
$\forall A\ B\ C\ D : \Point,\ \rk(\{A, B, C, D\}) \leq 4$
\item[$(A_{10})$]
$\forall\ A\ B\ C\ A'\ B'\ : \Point \exists\ M : \Point, 
  \rk(\{A,B,C\}) = 3\ \wedge\ \rk(\{A',B'\}) = 2 \Rightarrow \rk(\{A,B,C,M\}) = 3\ \wedge\ \rk(\{A',B', M\}) = 2 $
\end{itemize}
\caption[]{Rank axiom system for projective 3D-space}\label{rps}
\end{table}

Let us treat a rather simple example in 3D in order to illustrate how the axioms of such a rank function can be used to prove incidence projective theorems. We will not go into details because the proofs can become very long.

\begin{theorem}[Intersection of two planes]\label{thm:2planes}
In a 3D incidence projective space, the intersection of two different planes is a line.
\end{theorem}

\begin{figure}[h]
\begin{center}
  \includegraphics[scale=0.3]{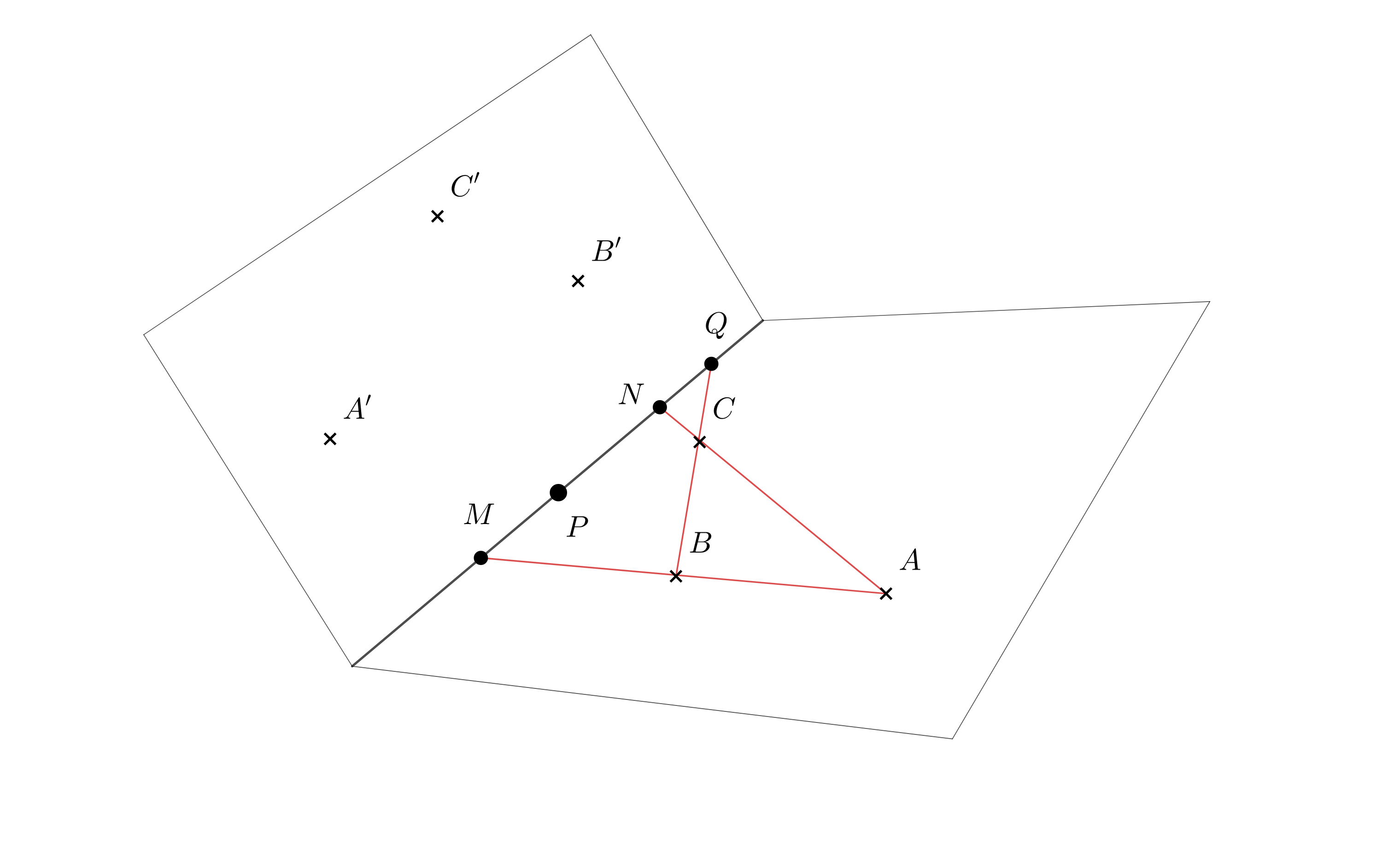}
 \caption{Intersection of two planes}\label{Fig:inter2planes}
\end{center}
\end{figure}

\vspace{-0.1cm}
We first translate this statements in terms of ranks:
\smallskip
\\
$
\forall\ A\ B\ C\ : \Point\ \forall\ A'\ B'\ C' : \Point\ \exists\ M\ N\ : \Point\ \forall\ P : \Point,\\
\rk(\{A,B,C\}) = 3\ \wedge\ \rk(\{A',B',C'\})=3\ \wedge\ \rk({A, B', C', A', B', C'})=4 \Rightarrow\\
\rk(\{A,B,C,M\}) = 3\ \wedge\ \rk(\{A',B',C',M\})=3 \wedge \\
\rk(\{A,B,C,N\}) = 3\ \wedge\ \rk(\{A',B',C',N\})=3 \wedge \rk(\{M,N\}) =2 \wedge \\
( \rk(\{M,N,P\})=2 \Leftrightarrow \rk(\{A,B,C,P\}) = 3\ \wedge\ \rk(\{A',B',C',P\})=3)
$
\smallskip

\noindent In natural language, points $A,B,C$ and points $A', B', C'$ define two different planes (they are different because of $\rk({A, B', C', A', B', C'})=4$). Then, there are two different points $M$ and $N$ defining a line which is the intersection of the two planes.
We start by proving that, there is at least a point in the intersection:

\begin{lemma}
 With the previous notations, there is at least one point $M$ in the intersection of the two planes.
\end{lemma}

It is clear that $\rk(\{A,B\})=2$ (by axiom $(A_3)$: $\rk(\{A,B,C\}) + \rk(\emptyset) \leq \rk(\{A,B\}) + \rk(\{C\})$). Axiom $(A_{10})$ with points $A',B',C'$ for the plane and $A,B$ for the line, introduces $M$ such that $\rk(\{A,B,M\})=2 \wedge \rk(\{A', B', C', M\})=3$.
Now, with ($A_3$), we have: 
$$
\rk(\{A, B, C, M\}) + \rk(\{A,B\}) \leq \rk(\{A,B,C\}) + \rk(\{A,B,M\}) \mbox{\hspace{1cm}} (1)
$$
Thus, $\rk(\{A,B,C,M\}) \leq 3$, and since $\{A,B,C\} \subseteq \{A,B,C,M\}$, we have $\rk(\{A,B,C,M\}) = 3$: $M$, that is $M$ belongs to plane defined by $A, B$ and $C$.

In the same way, we prove the existence of two points $N$ and $Q$ such that $\rk(\{A,C,N\})=2$ and $\rk(\{A', B', C', N\})=3$ for $N$, and ,$\rk(\{B,C,Q\})=2$ and $\rk(\{A', B', C', Q\})=3$ for $Q$.\\
Thus, we get three points $M, N$ and $Q$ in the intersection of the two planes. And, we have now to prove that among these 3 points, two are different, that is:

\begin{lemma}
 In an incidence projective plane, if three points $M, N$ and $Q$ are on the three edges of a triangle $ABC$, then at least two of these three points are different.
\end{lemma}

Using the matroid syntax, we have:\\
$
\forall\ M\ N\ Q\ A\ B\ C : \Point\ \rk(\{A,B,C\}) = 3\\
\wedge\ \rk(\{A,B,M\}) = 2 \ \wedge\ \rk(\{A,C,N\}) = 2\ \wedge\ \rk(\{B,C, Q\}) = 2\ \Rightarrow \\
\rk(M,N,Q) \geq 2
$

This lemma is interesting since we do not want to prove an exact value of the rank, but just a lower bound.
Using twice the method used to prove relation (1), it is easy to prove that $\rk(\{A,B,C,M,N,Q\})=3$.
We face then to a choice: considering geometry it is evident that there are two cases depending on whether $A=M$ or $A\neq M$ (we could also consider $B$ and $M$ or $A$ and $N$ or $C$ and $N$, \ldots).\\

\textit{Case }$A \neq M$. We have $\rk(\{A,M\})=2$.\\
Then $\rk(\{A,C,M,N,Q\})=3$ because of the inequalities 
\begin{itemize}
 \item $\rk(\{A,C,M,N,Q\}) \leq \rk(\{A,B,C,M,N,Q\})$ (by replacing \\
 $\rk(\{A,B,C,M,N,Q\})$, we get $\rk(\{A,C,M,N,Q\}) \leq 3)$
 \item $\rk(\{A,B,C,M,N,Q\}) + \rk(\{A,M\}) \leq \rk(\{A,B,M\}) + \rk(\{A,C,M,N,Q\})$ (that is 
 $3 + 2 \leq 2 + \rk(\{A,C,M,N,Q\})$)
\end{itemize}

Finally, considering the inequality:
$$\rk(\{A,B,C,M,N,Q\}) + \rk(\{N\}) \leq \rk(\{M,N,Q\}) + \rk(\{A,C,N\})$$
we have $\rk(\{M,N,Q\}) \geq 2 $.\\

\textit{Case }$A = M$. We have $\rk(\{A,M\})=1$.\\
By $\rk(\{A,B,C,M,N,Q\}) + \rk(\{M\}) \leq \rk(\{A, M\}) + \rk(\{B,C,M, N, Q\})$, we prove that $\rk(\{B,C,M, N, Q\})=3$ and we use this to prove $\rk(\{M,N,Q\}) \geq 2 $.\\
\emph{Qed}

By choosing two points between $(M,N,Q)$ we prove Theorem \ref{thm:2planes} first part: there are at least two different points in the intersection of the two planes.

This demonstrates that even a simple proof can be tedious. We will develop later the second part of the proof with the help of an automatic prover named Bip (matroid Based Incidence Prover)\footnote{\url{https://github.com/pascalschreck/MatroidIncidenceProver}}. Bip is not as smart as a modern IA can be, but nice enough to take in charge the tedious part of such proofs. We present this prover in the next section.

%---------------------------------------------------------------------------------
%
%       A matroid Based Incidence Prover
%
%---------------------------------------------------------------------------------

\section{Bip, a matroid Based Incidence Prover}\label{sect:Bip}

The reader may wonder about the appropriateness of using a rank function to make proofs in incidence geometry. From the previous example, this approach does not lead to straightforward proofs and seems far from geometric intuition for both the author and the reader. Actually the great advantage of this method is that it allows a certain amount of automation whatever the dimension.

During his PhD, David Braun implemented the kernel of a prover able to solve the matroid completion problem: given certain values of a rank function on a set $E$, deduce the value of the function for all the subsets of $E$\cite{Braun2019}. This solver exploits the axioms given above which do not contain existential quantifications, that is mainly axioms $(A_1)$ to $(A_4)$. The solver is also able to yield a Coq proof from the history of the computation of the rank function.
Besides, the proof of equivalence between the geometric and the algebraic approaches has been proved by our team in a previous work (\cite{desargueslong} and \cite{BMS19}). This allows us to consider mechanizing theorems expressed in geometric terms, the constraints expressed through the rank function being in a way, the ``assembling language'' of the prover.
We added some features to this prover and we call it now Bip.

We only give a brief description here, the interested reader can consult David's thesis \cite{Braun2019} (in French), or another brief introduction in \cite{BMS21}. 
%If he is patient enough, he can also wait for the publication of an article in preparation . 
The algorithm is based on a graph whose nodes are labeled by a set, a
lower bound and an upper bound for the value of the rank function for
this set. Initially, the whole powerset is represented. Then the rules
(see Table \ref{tab:ps}) derived from the axioms $(A_1)$ to $(A_3)$, are used to narrow the interval between the minimal and the maximal rank. It is a brute force algorithm and its complexity is exponential.

\begin{table}[h]
\begin{enumerate}[label=\textbf{(RS\arabic*)},leftmargin=1.9cm]
\item \textbf{\si} $X$ $\subseteq$ $Y$ \et $rkMin(X)$ $>$ $rkMin(Y)$ \textbf{\alors} $rkMin(Y)$ $\leftarrow$ $rkMin(X)$ \label{itm:rs1}
\item \textbf{\si} $Y$ $\subseteq$ $X$ \et $rkMin(Y)$ $>$ $rkMin(X)$ \textbf{\alors} $rkMin(X)$ $\leftarrow$ $rkMin(Y)$ \label{itm:rs2}
\item \textbf{\si} $X$ $\subseteq$ $Y$ \et $rkMax(Y)$ $<$ $rkMax(X)$ \textbf{\alors} $rkMax(X)$ $\leftarrow$ $rkMax(Y)$ \label{itm:rs3}
\item \textbf{\si} $Y$ $\subseteq$ $X$ \et $rkMax(X)$ $<$ $rkMax(Y)$ \textbf{\alors} $rkMax(Y)$ $\leftarrow$ $rkMax(X)$ \label{itm:rs4}
\item \textbf{\si} $rkMax(X) + rkMax(Y) - rkMin(X \cap Y)$ $<$ $rkMax(X \cup Y)$\\ \textbf{\alors} $rkMax(X \cup Y)$ $\leftarrow$ ($rkMax(X) + rkMax(Y) - rkMin(X \cap Y)$) \label{itm:rs5} 
\item \textbf{\si} $rkMax(X) + rkMax(Y) - rkMin(X \cup Y)$ $<$ $rkMax(X \cap Y)$\\ \textbf{\alors} $rkMax(X \cap Y)$ $\leftarrow$ ($rkMax(X) + rkMax(Y) - rkMin(X \cup Y)$) \label{itm:rs6}
\item \textbf{\si} $rkMin(X \cap Y) + rkMin(X \cup Y) - rkMax(Y)$ $>$ $rkMin(X)$\\ \textbf{\alors} $rkMin(X)$ $\leftarrow$ ($rkMin(X \cap Y) + rkMin(X \cup Y) - rkMax(Y)$) \label{itm:rs7} 
\item \textbf{\si} $rkMin(X \cap Y) + rkMin(X \cup Y) - rkMax(X)$ $>$ $rkMin(Y)$\\ \textbf{\alors} $rkMin(Y)$ $\leftarrow$ ($rkMin(X \cap Y) + rkMin(X \cup Y) - rkMax(X)$) \label{itm:rs8}
\end{enumerate}
\caption[Rules used until saturation]{Rules used until saturation ($\leftarrow$ means assignation).}\label{tab:ps}
\end{table}

\begin{algorithm}[h]
 \SetKwInOut{Input}{Input(s)}\SetKwInOut{Output}{Output(s)}
 \Input{Initialized powerset of the initial set of points}
 \Output{Powerset updated according to rules (RS1) to (RS8)}
 \textbf{While} a rank information was modified in the last step \textbf{Do}\\
 \hspace*{0.5cm} \textbf{ForEach} subset $X$ of $E$ \textbf{Do}\\
 \hspace*{1cm} \textbf{ForEach} subset $Y$ of $E$ such that $X$ $\neq$ $Y$ \textbf{Do}\\
 \hspace*{1.5cm} \textbf{ForEach} rewriting rule (RSi) $i\in [1;8]$ \textbf{Do}\\
 \hspace*{2cm} \textbf{If} the property (PSi with $i\in [1;8]$) is not verified \textbf{Then}\\
 \hspace*{2.5cm} apply the corresponding rewriting rule which updates \\ 
 \hspace*{2.5cm} the rank interval to ensure $RkMin$ $<$ $RkMax$\\
 \hspace*{2cm} \textbf{End If}\\
 \hspace*{1.5cm} \textbf{End ForEach}\\
 \hspace*{1cm} \textbf{End ForEach}\\
 \hspace*{0.5cm} \textbf{End ForEach}\\
 \textbf{End While}
 \caption[Saturation step]{Saturation step.}\label{alg:sat1}
\end{algorithm}

The algorithm simply consists in the following steps:
\begin{itemize}
 \item choosing two sets through two nested loops,
 \item computing their intersection and union if needed,
 \item applying a rule from Table \ref{tab:ps} on these sets
\end{itemize}
until a contradiction appears or nothing changes anymore. 
Throughout this process, the prover records the application of each rule. Then the trace of the deductions leading to the computation of a new rank for a given set of points is rebuilt. The trace is then extracted as a certificate and fed to the Coq proof assistant which verifies and validates it.

This version was aiming to help a mathematician in proving theorems in incidence projective geometry by handling small but tedious parts of a proof. But as said before the existential quantification is not taken into account. 
 It is a serious limitation because it is common to add auxiliary points in order
 to prove theorems.

Considering the initial target, the solver produced a monolithic proof of a single lemma with a rank equality as conclusion. 
This was not critical since the number of points was small and the proofs not very long. But we want to check its ability to discover long proofs on more complicated theorems. Therefore, we modify it so that it is more independent of the dimension, more robust and more easily usable.

We modify this prover by mainly adding:
\begin{itemize}
\item A change in the data structures in such a way that different dimensions can be taken into account. Now dimensions 4 and 5 can be considered.
 \item A text interface to facilitate communication with Coq or a human user: this allows one to write a statement without having to hard code it within bit-fields in language C. To this end, we design a simple ad-hoc language able to describe incidence statements with ranks in a given dimension. 
 \item A systematic decomposition into simple lemmas instead of building a monolithic proof. The whole proof of a theorem will be a bit longer but Bip is now able to handle huge proofs with more than 100,000 lines. This also allows us to consider an interactive and incremental use of the prover.
 \item The possibility of having several conclusions per theorem as it often happens in dimension greater than 2. In this context, the systematic decomposition in lemmas helps to limit the size of the proof.
\end{itemize}

In the next section, we present some examples of theorems formally proved with the help of Bip.

%---------------------------------------------------------------------------------
%
%       Various examples
%
%---------------------------------------------------------------------------------

\section{Various examples}\label{sect:examples}

The following examples show that our prover is now able to handle the combinatorial aspects of proofs in higher dimensions.

\subsection{Intersection of two planes in 3D}
We resume the example started in Section~\ref{sect:matroid}. Points
$ABC$ and $A'B'C'$ define two distinct planes and we already proved
that there are at least two distinct points $M$ and $N$ in their
intersection. We now prove that the intersection of the two planes is a line by double inclusion. This time each proof is automatically built.
We first prove that the intersection is included in the line spanned by $M$ and $N$ (in short line $(MN)$) (See Table \ref{Tab:inter1} left \footnote{the syntax is a bit simplified, the files with the full statements can be find here: \url{https://github.com/pascalschreck/ADG2021}}).
The two planes are different (i.e. $rk(A,B,C, A', B', C') = 4$), and the six points span the whole space. 

\begin{table}[h]
\begin{minipage}{0.49\linewidth}
 \begin{verbatim}
points
  A B C A' B' C' M N P  
hypotheses
  A B C : 3         
  A' B' C' : 3       
  A B C A' B' C' : 4 
  M A B C : 3
  N A B C : 3
  P A B C : 3
  M A' B' C' : 3 
  N A' B' C' : 3
  P A' B' C' : 3
  M N : 2   
conclusion
  M N P : 2  # collinearity
    
 \end{verbatim}
\end{minipage} 
\begin{minipage}{0.49\linewidth}
 \begin{verbatim}
   points
     A B C A' B' C' M N P 
   hypotheses
     A B C : 3      
     A' B' C' : 3       
     A B C A' B' C' : 4    
     M A B C : 3        
     N A B C : 3
     P A B C : 3
     M A' B' C' : 3       
     N A' B' C' : 3
     P A' B' C' : 3
     M N : 2              
     M N P : 2             
   conclusion       
     A B C P : 3    # P is in (ABC)
     A' B' C' P : 3 
 \end{verbatim}
\end{minipage}
\caption{In 3D, the intersection of two planes is included into a line (right), and line $(MN)$ is included in the intersection (left).}\label{Tab:inter1}\label{Tab:inter2}
\end{table}

The size of the output files is very dependent on the number of points in the statement. With 9 points, the size of the file with the whole rank function is about 58 Kb, and the size of the proof in Coq is about 43 Kb (and a bit less than 500 lines). 
Generating the proof takes less than 1s and its verification by Coq
only a few seconds.
Now, we prove that $(MN) \subset (ABC) \cap (A'B'C')$. The statement is very similar (See Table \ref{Tab:inter2} right)
apart from the fact that the conclusion is now the conjunction of two equalities. In this case, Bip yields a Coq proof which includes, among others, a final theorem with the conjunction of the two terms in conclusion. The size of the produced files are equivalent to the previous theorem.

\subsection{Dandelin-Gallucci's theorem}

Dandelin-Gallucci's \emph{theorem} %then
establishes a strong link
between an iconic 2D property (Pappus) and a truly 3-dimensional one (Dandelin-Gallucci):
%We can now state Dandelin-Gallucci's \emph{theorem}:
\begin{theorem}[Dandelin-Gallucci]
 In a projective incidence space whose dimension is greater than or
 equal to 3, Dandelin-Gallucci's property and Pappus's property are equivalent.
\end{theorem}

This proof only uses basic knowledge on incidence
geometry. Fig. \ref{Fig:DG1} illustrates the configuration and names
some interesting lines. 
This sketch highlights the role of Pappus's points $X$, $Y$ and $Z$
which are not part of the initial configuration and are later used to construct the point $R$ as the intersection of lines
$(YM)$ and $(ZN)$. We then show that this new point $R$ is also the
intersection of lines $d$ and $h$.  
The details of proof can be found in Horv\'{a}th's article~\cite{galluccibis}. 

\subsubsection{From Pappus to Dandelin-Gallucci}

As mentioned above, points $X$, $Y$ and $Z$ materialize an instance of
Pappus's property. This instance is chosen by hand. There are many
possibilities among such configurations and considering each of them
would dramatically increase the computation time and space leading to
unmanageable proofs.  

\begin{figure}[h]
\begin{center}
  \includegraphics[scale=0.37]{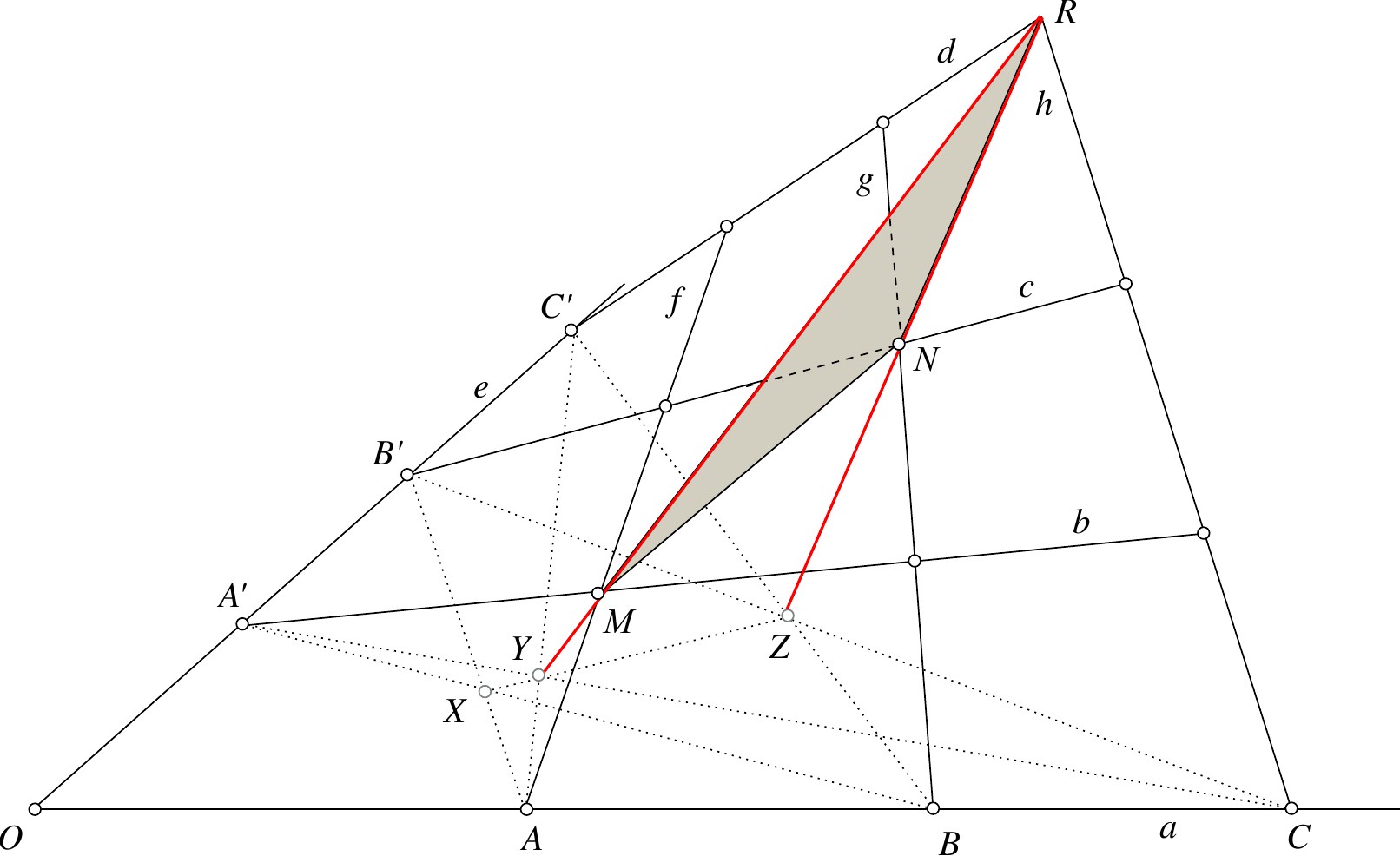}\\
\end{center}
 \caption{ From Pappus to Dandelin-Gallucci \label{Fig:DG1} (lines are noted with lower case letters $a$, $b$, \ldots). 
 }
\end{figure}

This triple of points being defined, we add their collinearity as an hypothesis. This could be done by a rule of the solver, and actually, such a rule was implemented but on the one hand finding a complicated pattern  is a new source of complexity and on the other hand, it seems very artificial to automatically retrieve what was added by hand.

The point $R$ is then defined as the intersection of lines $(YM)$ and
$(ZN)$ (See Figure \ref{Fig:DG1}), but the collinearity of these lines
has to be proved before. This is why there are two parts of the
proof.  We do not give the two statements here, but all the examples can be found on the Git repository already mentioned. 
The whole figure contains 19 points but only 17 of them are required giving lighter outputs: the size of the rank function file is about 16.4 Mb and the Coq proof about 1.7 Mb (2200 lines). The computation time of the second part is about half an hour.

\subsubsection{From Dandelin-Gallucci to Pappus}

The opposite direction of the proof is similar but a lot of points have to be added by hand:  10 points are involved by the hypotheses and 17 points are needed by the proof. These points all come from the construction of a Dandelin-Gallucci configuration which is done by hand following the Horv{\'a}th's article \cite{galluccibis}.
With 17 points, the size of the rank function is the same as before, but the Coq proof is a bit longer with about 50,000 lines.

\subsection{Desargues's theorem(s) in $n$D}
\begin{figure}[h]
\begin{center}
\includegraphics[scale=0.25]{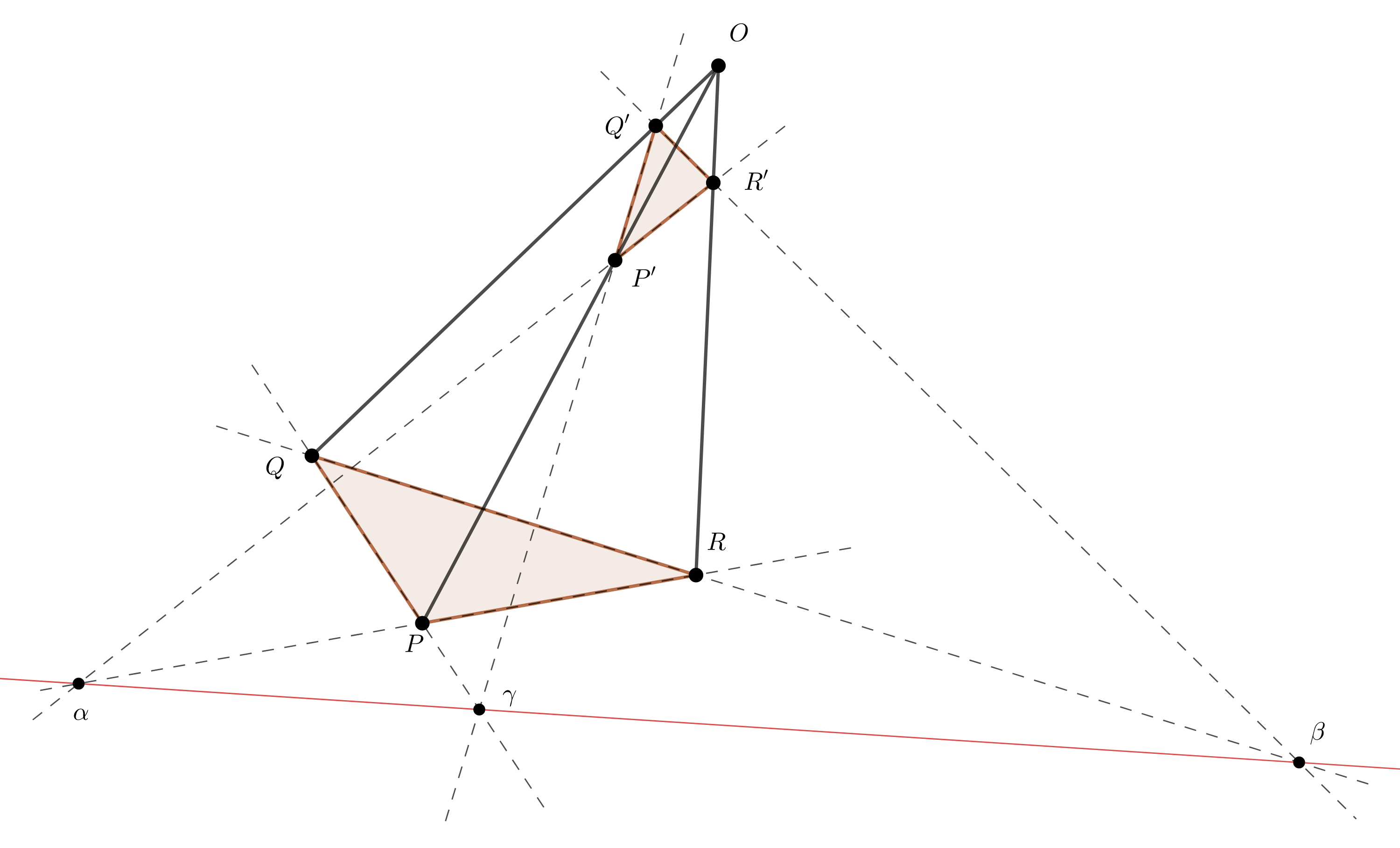}
\end{center}
\caption[Desargues' theorem \ref{thm:desargues}, in its 3D configuration]{Desargues' theorem \ref{thm:desargues}, in its 3D configuration}\label{fig:desargues3D}
\end{figure}

We recall Desargues's property in section~\ref{sect:geometry} as a very important property in projective incidence geometry. It does not hold in general in a 2D plane, for example in the Moulton's plane \cite{moulton}, but it becomes a theorem when the considered plane is embedded in a 3D space. This theorem was proved interactively using Coq a few years ago \cite{desargueslong}. It proceeds as follows: we prove the property
in the case where the 2 triangles are not coplanar
\footnote{This configuration was improperly  called Desargues 3D.} (as shown in Fig. \ref{fig:desargues3D}). Then, one can derive the final proof in the case where the 2 triangles are coplanar. The two proofs are now automatically produced by Bip. 

With 15 points, it takes less than 2 minutes to solve this statement. The file which contains the rank function has a size of 4 Mb (a bit more than 32 * 118 Kb) and the Coq file is about 1.0 Mb, that is a bit less than 13000 lines.

%-------------------------------------------------------------------------------------
%
% Higher dimensions
%
%-------------------------------------------------------------------------------------

\subsubsection{Higher dimensions}
The Desargues's theorem has a very combinatorial nature and it has in fact a version in any dimension greater than 2.
%\iffalse
\begin{figure}
 \begin{center}
  \includegraphics[scale=0.18]{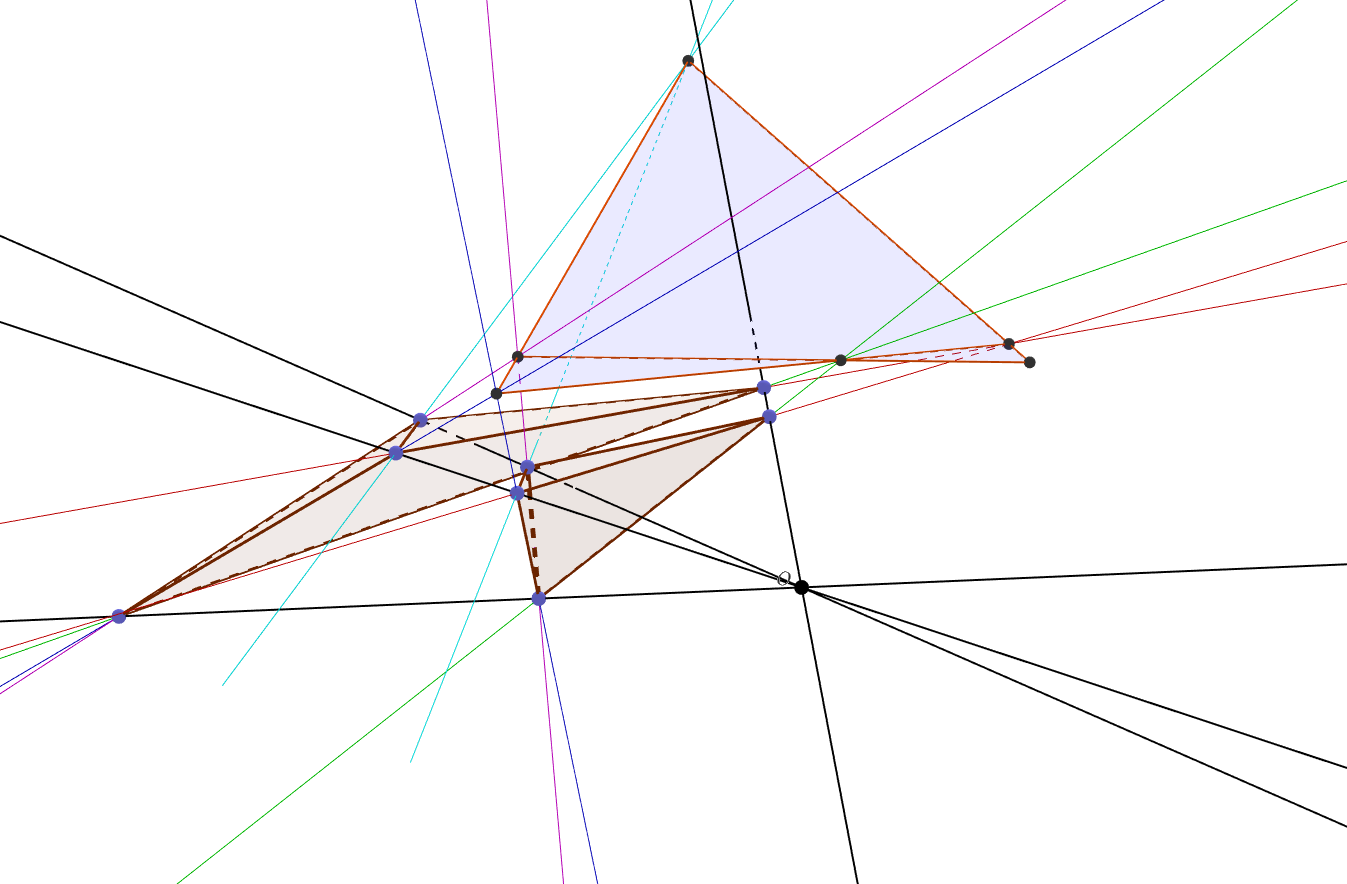}
 \end{center}
\caption{In 3D: the tetrahedrons are in brown and the complete quadrilateral in blue.}\label{Fig:realD3D}
\end{figure}
%\fi
For instance in dimension 3, it states that given two tetrahedrons $T$ and $T'$ which are in perspective from a point $O$, the 6 points defined by the intersection of the corresponding edges of $T$ and $T'$ are in a plane and form a complete quadrilateral 
%-------------------------------------figure mise en commentaire
(See Figure \ref{Fig:realD3D}). 
15 points are involved in this theorem but it is solved in a few minutes. The Coq proof is about 31000 lines long.

We also succeed in proving the 4D version of this theorem: given 2 pentachores $P$ and $P'$ which are in perspective from a point $O$, the 10 points defined by the intersection of the corresponding edges define a 3D space and form a figure which is precisely the one depicted on Figure \ref{fig:desargues3D}. 21 points are involved in this theorem, and it takes about one week to solve it.

In dimension $n$, let us call a $n$-complete hyper-tetrahedron the configuration $G$ of $\frac{(n+1)(n+2)}{2}$ points and defined by $G = H \cup H'$ where
\begin{itemize}
 \item $H$ is a set of $n+1$ points in general position, that is a hypertetrahedron in dimension $n$,
 \item $H'$ is the intersection of the hypertetrahedron by an independent hyperplane $P$; in other words, $H'$ is a set of ${n+1} \choose 2$ new points each of them is the intersection of an edge of $H$ with hyperplane $P$.
\end{itemize}
A 1-complete hypertetrahedron consists of 3 collinear points.
A 2-complete hypertetrahedron is a complete quadrilateral and a 3-complete hypertetrahedron is the configuration depicted at Figure \ref{fig:desargues3D}. We have then

\begin{theorem}[Desargues's theorem in dimension $n+1$]
 Let $E$ an incidence space of dimension $n+1$ and 2 two hypertetrahedrons $T$ and $T'$ in this space in perspective from a point $O$ independent of each tetrahedron, that is, $O$ does not belong to any hyperface of $T$ or $T'$. Then the intersection points of the corresponding edges of $T$ and $T'$ form a $n$-complete hypertetrahedron.
\end{theorem}

The formal proof of this theorem is out of reach of our prover since it requires a proof by recursion on the dimension (See \cite{Bell} for an Euclidean point of view).

\subsection{Simple examples in dimension 4 and 5}

In dimension 4 and 5, we tested Bip with simple examples around the intersection of two flats. For instance, in 4D, the intersection of two different planes is a point (See the Git repository at the URL given above for more examples).

We present here the statements expressing that the intersection of two different hyperplanes is a 3D space modulo the existence of the four different points $I, J, K, L$, see Tables \ref{table:5d} and \ref{table:5d_2}.

\begin{table}[ht]
 \begin{minipage}{0.55\linewidth}
 \begin{verbatim}
points 
  A B C D E Ap Bp 
  Cp Dp Ep I J K L M
hypotheses
  A B C D E : 5
  Ap Bp Cp Dp Ep : 5
  A B C D E Ap Bp Cp Dp Ep : 6
  I A B C D E : 5
  J A B C D E : 5
  K A B C D E : 5
  L A B C D E : 5
  M A B C D E : 5
\end{verbatim}
\end{minipage}
\begin{minipage}{0.45\linewidth}
\begin{verbatim}
    I Ap Bp Cp Dp Ep : 5
    J Ap Bp Cp Dp Ep : 5
    K Ap Bp Cp Dp Ep : 5
    L Ap Bp Cp Dp Ep : 5
    M Ap Bp Cp Dp Ep : 5
    I J K L : 4
    I J K M : 4
    I J L M : 4
    I K L M : 4
    J K L M : 4
  conclusion
    I J K L M : 4
    
\end{verbatim}
\end{minipage}
\caption{The intersection of 2 hyperplanes in a 5D space is included in a 3D-space}\label{table:5d}
\end{table}

\begin{table}[ht]
 \begin{minipage}{0.55\linewidth}
 \begin{verbatim}
points 
  A B C D E Ap Bp Cp Dp Ep I J K L M
hypotheses
  A B C D E : 5
  Ap Bp Cp Dp Ep : 5
  A B C D E Ap Bp Cp Dp Ep : 6
  I A B C D E : 5
  J A B C D E : 5
  K A B C D E : 5
  L A B C D E : 5
  
\end{verbatim}
\end{minipage}
\begin{minipage}{0.45\linewidth}
\begin{verbatim}
    I Ap Bp Cp Dp Ep : 5
    J Ap Bp Cp Dp Ep : 5
    K Ap Bp Cp Dp Ep : 5
    L Ap Bp Cp Dp Ep : 5
    I J K L : 4
    I J K L M : 4
  conclusion
    M A B C D E : 5
    M Ap Bp Cp Dp Ep : 5
\end{verbatim}
\end{minipage}
\caption{A 3d-space is included in the intersection of 2 hyperplanes in a 5D space.}\label{table:5d_2}
\end{table}

%---------------------------------------------------------------------------------
%
%       Conclusion
%
%---------------------------------------------------------------------------------
\section{Conclusion}
\label{sect:conclusion}
In this paper, we propose an extension to our automatic prover Bip, so that it can deal with statements in higher dimension ($\geq 3$). We summarize the usual way the prover works by revisiting Dandelin-Gallucci’s theorem in 3D. We then present some preliminary results about some theorems in dimension 4 and also in dimension 5. Our prover can easily deal with these simple examples, even if proving statements in higher dimensions requires running the prover longer.  Overall these first experiments show that our prover scales well to deal with higher dimension statements. 

We plan to pursue our investigations and to extend Bip in several directions.
In particular, we want to improve the core mechanism of Bip in order to have a better complexity and to tackle for instance the reciprocal theorem of Desargues in 4D and Desargues's theorem and its reciprocal in 5D.
We also plan to specify a better language to express theorems with matroid terms in order to improve its expressiveness  and readability.

\ifx\append\undefined
   \bibliographystyle{eptcs}
   \bibliography{biblio}
   \end{document}
\fi